# Fast Updates on Read-Optimized Databases Using Multi-Core CPUs

Jens Krueger[†], Changkyu Kim[⋆], Martin Grund[†], Nadathur Satish[⋆], David Schwalb[†],
Jatin Chhugani[⋆], Hasso Plattner[†], Pradeep Dubey[⋆], Alexander Zeier[†]

[†]Hasso-Plattner-Institute, Potsdam, Germany  
Contact: jens.krueger@hpi.uni-potsdam.de

[⋆]Parallel Computing Lab, Intel Corporation  
Contact: changkyu.kim@intel.com

## ABSTRACT

Read-optimized columnar databases use differential updates to handle writes by maintaining a separate write-optimized delta partition which is periodically merged with the read-optimized and compressed main partition. This merge process introduces significant overheads and unacceptable downtimes in update intensive systems, aspiring to combine transactional and analytical workloads into one system.

In the first part of the paper, we report data analyses of 12 SAP Business Suite customer systems. In the second half, we present an optimized merge process reducing the merge overhead of current systems by a factor of 30. Our linear-time merge algorithm exploits the underlying high compute and bandwidth resources of modern multi-core CPUs with architecture-aware optimizations and efficient parallelization. This enables compressed in-memory column stores to handle the transactional update rate required by enterprise applications, while keeping properties of read-optimized databases for analytic-style queries.

## 1. INTRODUCTION

Traditional read-optimized databases often use a compressed column oriented approach to store data [25, 18, 28]. Performing single inserts in such a compressed persistence can be as complex as inserting into a sorted list [11]. One approach to handle updates in a compressed storage is a technique called differential updates, maintaining a write-optimized delta partition that accumulates all data changes. Periodically, this delta partition is combined with the read-optimized main partition. We refer to this process as *merge* throughout the paper, also referred to as checkpointing by others [11]. This process involves uncompressing the compressed main partition, merging the delta and main partitions and recompressing the resulting main partition. In contrast to existing approaches, the complete process is required to be executed during regular system load without downtime. The update performance of such a system is limited by two factors — (a) the insert rate for the write-optimized structure and (b) the speed with which the system can merge the accumulated updates back into the read-optimized partition. Inserting into the write-optimized structure can be performed fast if the size of the structure is kept small enough. As an additional benefit, this also ensures that the read performance does not degrade significantly. However, keeping this size small implies merging frequently, which increases the overhead of updates.

To the best of our knowledge we are not aware of any sophisticated implementation and therefore compare against a naïve implementation. Based on the result of analyzing 12 SAP Business Suite customer systems, we found that current systems would merge approx. 20 hours every month, while supporting a maximum of ~1,000 updates per second (see Section 2 for more detail). In read-mostly scenarios this limitation is not a major problem since the workload can be stopped during reload, modifications are invisible until applied in batch or performance degradation is acceptable. However, when engineering a system for both transactional and analytical workloads as described in [22, 17, 13], it becomes essential to reduce the merge overhead and to support the required single update rates for handling transactional workloads. Systems under load will have to cope with even longer times for merging or face considerable downtimes every month. This causes a scheduling problem, which is particularly critical when considering fully utilized systems of globally active enterprises where downtimes are not acceptable. Additionally, growing data volumes over time and expanding amounts of captured data in the transactional processes further intensify the impact of the merge process.

*Contributions.* We propose an optimized online merge algorithm for dictionary encoded in-memory column stores, enabling them to support the update performance required to run enterprise application workloads on read-optimized databases. More specifically, we make the following contributions:

(i) We make design choices based on analyses of real enterprise systems.

(ii) We develop multi-core aware optimizations taking underlying CPU caches and thread-level parallelism into account, resulting in a speed-up of 30 times over unoptimized serial code.

(iii) We verify our analytical model in terms of compute and memory bandwidth requirements for current and applicable for upcoming architectures.

The remainder of the paper is structured as follows. First, Section 2 presents data analyses of the most influential parameters for the merge process on real enterprise systems, followed by Section 3 with an overview of our system. Then, 4 and 5 introduce our merge algorithm. Section 6 describes an optimized and parallelized version which is closely evaluated in Section 7. Related work is discussed in Section 8 and Sections 9 and 10 conclude the paper.

## 2. ENTERPRISE DATA CHARACTERISTICS

This section motivates combined transactional and analytical systems using a compressed in-memory column store. We base our findings on a thoroughly conducted analysis of SAP Business Suite customer systems. In total we analyzed customer data of 12 companies with 74.000 tables per installation.[1] We analyzed the application behavior, the system's workloads, table sizes, distinct values and current merge durations in order to derive scenarios to validate our optimized merge implementation.

*Applications.* Traditionally, the database market divides into transaction processing (OLTP) and analytical processing (OLAP) workloads. OLTP workloads are characterized by a mix of reads and writes to a few rows at a time, typically through a B+Tree or other index structures. Conversely, OLAP applications are characterized by bulk updates and large sequential scans spanning few columns but many rows of the database, for example to compute aggregate values. Typically, those two workloads are supported by two different types of database systems – transaction processing systems and warehousing systems.

In contrast to this classification, single applications such as Demand Planning or Available-To-Promise exist, which cannot be exclusively referred to one or the other workload category. These applications initiate a mixed workload in terms of that they process small sets of transactional data at a time including write operations and simple read queries as well as complex, unpredictable mostly read operations on large sets of data with a projectivity on a few columns only. Furthermore, there is an increasing demand for "real-time analytics" – that is, up-to-the moment reporting on business processes that have traditionally been handled by data warehouse systems. Although warehouse vendors are doing as much as possible to improve response times (e.g. by reducing load times), explicit separation between transaction processing and analytics systems introduces a fundamental bottleneck in analytical scenarios. Predefinition of data that is extracted, transformed and loaded into the analytics system leads to the fact that analytics-based decision are made on a subset of potential information. This separation of systems prevents transactional applications from using analytical functionality throughout the transaction processing due to the latency that is inherent in the data transfer.

*Workload.* As a follow-up of the previous section, we analyzed the enterprise application workload as of today. Figure 1 shows the query distribution of key lookups, table scans, range selects, inserts, modifications and deletes for the transactional and analytical systems . In total, more than 80% of all queries are read access — for OLAP systems even over 90%. While this is the expected result for analytical systems, the high amount of read queries on transactional systems is surprising as this is not the case in traditional workload definitions. Consequently, the query distribution leads to the concept of using a read-optimized database for both transactional and analytical systems. Even though most of the workload is read-oriented, ~17% (OLTP) and ~7% (OLAP) of all queries are updates. A read-optimized database supporting both workloads has to be able to support this amount of update operations. Additional analyses on the data have shown an update rate varying from 3,000 to 18,000 updates/second. In contrast, the TPC-C benchmark, which has been the foundation for optimizations over the last decade, has a higher write ratio (46%) compared to our analysis (17%).

---

[1] The number of tables is constant per installation, but differently used based on the modules in use.

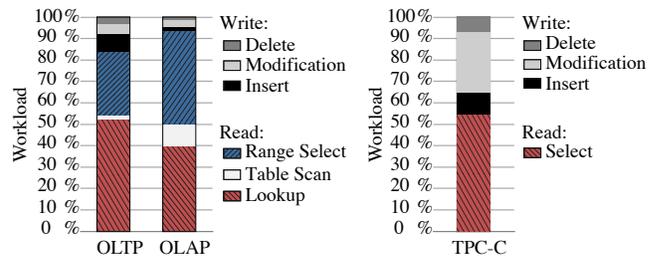

Figure 1: Distribution of query types extracted from customer database statistics, comparing OLTP and OLAP workloads. In contrast, the TPC-C benchmark has a higher write ratio.

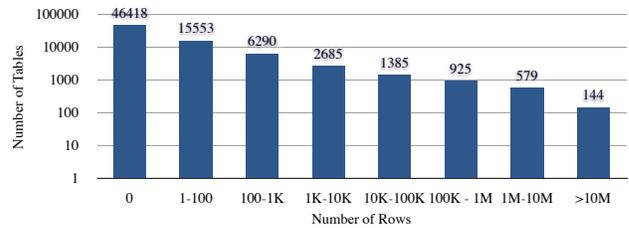

Figure 2: All 73,979 tables clustered by number of rows.

*Table Size.* Figure 2 presents the different sizes of the analyzed tables, while the tables are grouped by size. Despite the huge total amount of tables, only a limited number have a high amount of rows. Merging the smaller tables is easier to schedule due to their size and the fact that they will not be updated as often as the larger ones. An efficient merge process that provides the required execution speed will allow for fast updates on the most active tables without interfering with the system performance. Figure 3 shows a closeup of the 144 largest tables from Figure 2 with their distribution of the number of rows and columns, creating the major part of the merge costs of the system.

*Distinct Values.* In order to estimate the respective dictionary sizes when applying dictionary encoding to the table, we analyzed the 21 most active tables of each customer. In summary, over 32 billion records and more than 400 million distinct values were inspected. We observed that enterprise data works on a well known value domain of each column. Most of the columns in financial accounting and inventory management work with a very limited set of distinct values, as depicted in Figure 4. Since most of the columns come with a given configuration of values and free value entries are very rare, enterprise application data is suitable for compression techniques by exploiting redundancy within the data and knowledge about the data domain for optimal results. Furthermore, columns with a small number of distinct values and a large value size heavily profit from dictionary encoding.

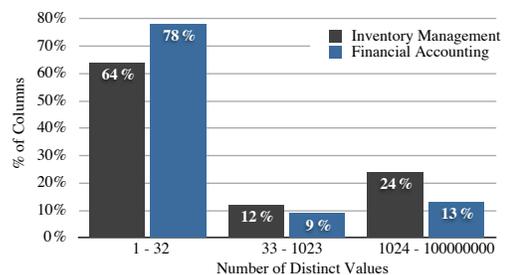

Figure 4: Distinct Values in Inventory Management and Financial Accounting.

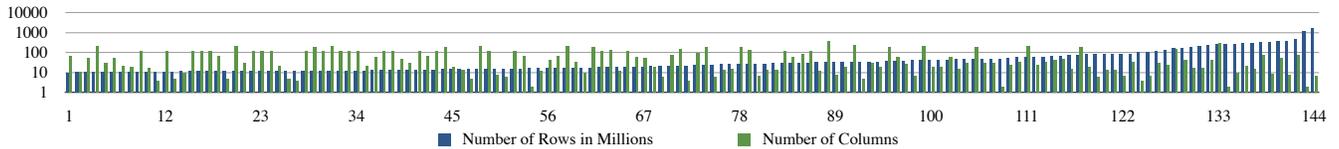

Figure 3: Overview of the 144 tables which have more than 10 million rows of one analyzed SAP Business Suite customer system. The tables are sorted by the number of rows and the abscissa shows the respective position. For every table, the number of rows (in millions) are displayed in blue and the number of columns in green.

*Merge Duration.* We concentrated on the 144 most intensively used tables. The number of rows varies from 10 million to 1.6 billion with an average of 65 million rows, whereas the number of columns varies from 2 to 399 with an average of 70. We picked the $VBAP$ table with sales order data of 3 years (33 million rows, 230 columns, 15 GB) and measured the merge of new sales order data from one month of 750,000 rows, taking 1.8 trillion CPU cycles or 12 minutes[2]. Converted, our initial implementation handled ~1,000 merged updates per second. Using this as an estimation for the complete system with a size of 1.5 TB, the total merge duration was around 20 hours every month.

*Summary.* Nowadays database management systems cannot fulfill the requirements of specific enterprise applications since they are either optimized for one or the other workload category, leading to a mismatch of enterprise applications regarding the underlying data management layer. Considering the trend of increasingly complex business processes, the new functionality provided by enterprise-specific data management, enabling ad-hoc queries on transactional events, will lead to an amplified usage of computing-intensive analytic-style queries. Based on the demand of analytics on up-to-date transactional data, the data characteristics of enterprise data as presented above and the query distribution as shown in Figure 1, we see a huge potential for read-optimized databases like in-memory column stores for becoming the main persistence of enterprise applications combining analytical and transactional workloads.

## 3. SYSTEM OVERVIEW

Based on the observations presented in the previous section, we present HYRISE [10]. HYRISE is an in-memory compressed vertical partition-able database engine. It is designed to support the update rate required by mixed workloads and to provide the high read performance of column stores. In this paper we concentrate on a fully decomposed storage layout. Tables are stored physically as collections of attributes and metadata. Each attribute consists of two partitions: main and delta partition.

The main partition is dictionary compressed — values in the tuples are replaced by encoded values from the dictionary. An ordered collection is used as a dictionary, allowing fast iterations over the tuples in sorted order. Additionally, the search operation can be implemented as binary search that has logarithmic complexity. As a result, most queries can be executed with a binary search (random access) in the dictionary while scanning the column (sequential access) for the encoded value only. This comes at a price: updates might change encoded values and require a rewrite of the complete partition. Consequently, direct updates in the main partition would be unacceptable slow.

In order to minimize the overhead of maintaining the sort order, incoming updates are accumulated in the write-optimized delta partition as described in [25]. In contrast to the main partition, data in the write-optimized delta partition is not compressed. In addition to the uncompressed values, a CSB+ tree [24] with all the unique uncompressed values of the delta partition is maintained per column. This approach provides acceptable insert performance and fast read performance with support for sorted iterations.

---
[2]On 2 x 6 core Intel® Xeon® processor X5650, 48 GB RAM, 2.6 GHz

The design decision of not compressing the delta partition has two important properties that affect the overall system performance. Firstly, memory consumption increases, leading to a potential drop in performance. Secondly, maintaining an uncompressed delta partition means that both read and update queries save extra random memory accesses for materialization. For a net performance benefit, the size of the delta partition must be kept small.

To ensure this, HYRISE executes a periodic merge process. In contrast to the definition of merging in [11], our understanding of the merge process is as follows: A merge process combines all data from the main partition as well as the delta partition to create a new main partition that then serves as the primary data store. To avoid confusion with merging query intermediate results, when talking about merging data in this paper we always refer to the merge process described above. The merge process is transactionally safe, as it works on a copy of the table and the merged table is committed atomically at the end. During the merge, incoming updates are stored in a temporary second delta, which becomes the primary delta when the merge result is committed. Interferences with other queries are minimized, as the table has to be locked only for a minimal period at the beginning and end of the merge. As a result, the merge only impacts other running queries if the system resources are greatly over-committed, causing resource contention. However, a scheduling algorithm can detect a good point in time to start and even pause and resume the merge process. We see two scheduling strategies: a) merging with all available resources and b) minimizing resource utilization by constantly merging in the background. The topics of transactional safety and scheduling of the merge process are both orthogonal to the topic of this paper. For the remainder, we assume that the merge uses all available resources.

In order to achieve high update rates, table modifications in HYRISE follow the insert-only approach. Therefore, updates are always modeled as new inserts and deletes only invalidate rows. We keep the insertion order of tuples and only the lastly inserted version is valid. Firstly, we chose this concept because only a small fraction of the workload are modifications as described in Section 2 and the insert-only approach allows queries to also work on the history of data. Secondly, we rejected to change the order of stored tuples for the benefit of better read performance in order to avoid surrogate tuple ids. In our system, the implicit offset of a tuple is always valid for all attributes of a table. Due to this design decision, columns cannot be sorted individually as in other column based systems, e.g. [25]. As memory bandwidth clearly is a bottleneck for our parallelized merge algorithm, we use dictionary encoding and bit-compression to reduce the transferred data from and to main memory in our prototype system. However, as the growth in CPU compute power exceeds the development of main memory bandwidth it could be desirable to apply further compression rou-

63

tines, trading less memory traffic for more CPU cycles. However, compression algorithms modifying the tuple order introduce dependencies between attributes and thus limit the scalability. Therefore, compression techniques like run length encoding are not applied in our prototype. We trade additional compression techniques for better update and merge performance, a simplified transaction handling and the possibility to store the change history for all records as it is required by enterprise systems.

*Terminology.* In order to avoid ambiguity, we define the following terms to be used in the rest of the paper.

1. Table: A relation table with $N_C$ columns, with one write (delta) and one read optimized (main) partition.
2. Update: Any modification operation on the table resulting in an entry in the delta partition.
3. Main Partition: Compressed and read-optimized part of the column.
4. Delta Partition: Uncompressed write-optimized part of the column where all updates are stored until the merge process is completed.
5. Value-Length: The number of bits allocated to store the uncompressed values for a given column.

## 4. UPDATE PERFORMANCE

In order to account for the time taken to perform the merging of the main and delta partitions for each column, we define the total time taken to perform updates as the

$$\text{Update Rate} = \frac{N_D}{T_U + T_M} \text{Updates/Second} \quad (1)$$

where $T_U$ defines the time taken to perform $N_D$ updates to the delta partitions of all columns and $T_M$ defines the time taken to perform the merging of the main and delta partitions of all columns of the table.

For supporting efficient updates, we want the update rate to be as high as possible. Furthermore, the update rate should also be greater than the minimum sustained update rate required by the specific application scenario. When the number of updates against the system durably exceed the supported update rate, the system will be rendered incapable of processing new inserts and other queries, leading to failure of the database system. In order to prevent this, we set a low target update rate of 3,000 and a high target update rate of 18,000 updates/second. Even though most of the workload issued is read-oriented this target is chosen based on the analyzed data and to cover various workloads.

An important performance parameter for our system is the frequency at which the merging of the partitions must be executed, as presented in Section 3. The frequency of executing the merging of partitions affects the size (number of tuples) of the delta partition. Computing the appropriate size of the delta partition before executing the merge operation is dictated by the following two conflicting choices:

(i) **Small delta partition** A small delta partition implies a relatively low overhead to the read query, implying a small reduction in the read performance. Furthermore, the insertion into the delta partition will also be fast. This means however that the merging step needs to be executed more frequently, thereby increasing the impact on the system.

(ii) **Large delta partition** A large delta partition implies that the merging is executed less frequently and therefore adds only a little overhead to the system. However, increasing the delta partition size implies a slower read performance due to the fact that the delta partition stores uncompressed values, which consume more compute resources and memory bandwidth, thereby appreciably slowing down read queries (scan, index lookup, etc.) Also, while comparing values in the main partition with those in delta partition, we need to look up the dictionary for the main partition to obtain the uncompressed value (forced materialization), thereby adding overhead to the read performance.

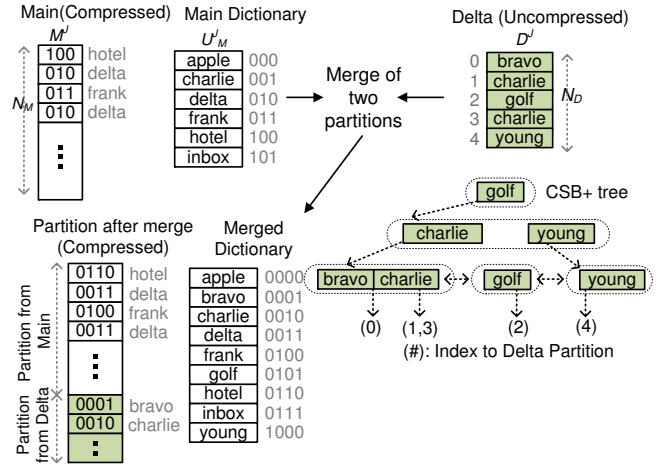

Figure 5: Example showing the data structures maintained for each column. The main partition is stored compressed together with the dictionary. The delta partition is stored uncompressed, along with the CSB+ tree. After the merging of the partitions, we obtain the concatenated compressed main column and the updated dictionary.

In our system, we trigger the merging of partitions when the number of tuples $N_D$ in the delta partition is greater than a certain pre-defined fraction of tuples in the main partition $N_M$.

### 4.1 Example: Merging Main and Delta

Figure 5 shows an example of a column with its main and delta partitions. Note that the other columns of the table would be stored in a similar fashion. The main partition has a dictionary consisting of its sorted unique values (6 in total). Hence, the encoded values are stored using 3 ( = $\lceil \log 6 \rceil$) bits. The uncompressed values (in gray) are not actually stored, but shown for illustration purpose. As described earlier in Section 3, the compressed value for a given value is its position in the dictionary, stored using the appropriate number of bits (in this case 3 bits).

The delta partition stores the uncompressed values themselves. In this example, there are five tuples with the shown uncompressed values. In addition, the CSB+ tree containing all the unique uncompressed values is maintained. Each value in the tree also stores a pointer to the list of tuple ids where the value was inserted. For example, the value "charlie" is inserted at positions 1 and 3. Upon insertion, the value is appended to the delta partition and the CSB+ tree is updated accordingly.

After the merging of the partitions has been performed, the main and delta partitions are concatenated and a new dictionary for the concatenated partition is created. Furthermore, the compressed values for each tuple are also updated. For example, the encoded value for "hotel" was 4 before merging and "6" after merging. Furthermore, it is possible that the number of bits that are required to store the compressed value will increase after merging. Since the number of unique values in this example is increased to 9 after merging, each compressed value is now stored using $\lceil \log 9 \rceil = 4$ bits.

| Description | Unit | Symbol |
|---|---|---|
| Number of columns in the table | - | $\mathbf{N}_C$ |
| Number of tuples in the main table | - | $\mathbf{N}_M$ |
| Number of tuples in the delta table | - | $\mathbf{N}_D$ |
| Number of tuples in the updated table | - | $\mathbf{N}'_M$ |
| For a given column $j; j \in [1 \ldots \mathbf{N}_C]$: | | |
| Main partition of the $j^{th}$ column | - | $\mathbf{M}^j$ |
| Merged column | - | $\mathbf{M}'^j$ |
| Sorted dictionary of the main partition | - | $\mathbf{U}_\mathbf{M}^j$ |
| Sorted dictionary of the delta partition | - | $\mathbf{U}_\mathbf{D}^j$ |
| Updated main dictionary | - | $\mathbf{U}_\mathbf{M}'^j$ |
| Delta partition of the $j^{th}$ column. | - | $\mathbf{D}^j$ |
| Uncompressed Value-Length | bytes | $\mathbf{E}^j$ |
| Compressed Value-Length | bits | $\mathbf{E}_C^j$ |
| Compressed Value-Length after merge | bits | $\mathbf{E}_C'^j$ |
| Fraction of unique values in delta | - | $\lambda_\mathbf{D}^j$ |
| Fraction of unique values in main | - | $\lambda_\mathbf{M}^j$ |
| Merge auxiliary structure for the main | - | $\mathbf{X}_\mathbf{M}^j$ |
| Merge auxiliary structure for the delta | - | $\mathbf{X}_\mathbf{D}^j$ |
| Cache Line size | bytes | L |
| Memory Traffic | bytes | MT |
| Number of available parallel threads | - | $\mathbf{N}_T$ |

Table 1: Symbol Definition. Entities annotated with *ı* represent the merged (updated) entry.

## 5. EFFICIENT MERGING ALGORITHM

We now describe the merge algorithm in detail and enhance the naïve merge implementation by applying optimizations known from join processing. Furthermore, we will parallelize our implementation and make it architecture-aware to achieve the best possible throughput. For the remainder of the paper we refer to the symbols explained in Table 1.

As explained in Section 3, we use a compression scheme wherein the unique values for each column are stored in a separate dictionary structure consisting of the uncompressed values stored in a sorted order. Hence, $|\mathbf{U}_\mathbf{M}^j| = \lambda_\mathbf{M}^j \cdot \mathbf{N}_M$ with $|X|$ denoting the number of elements in the set X. By definition, $\lambda_\mathbf{M}^j, \lambda_\mathbf{D}^j \in [0 \ldots 1]$. Furthermore, the compressed value stored for a given value is its index in the (sorted) dictionary structure, thereby requiring $\lceil \log |\mathbf{U}_\mathbf{M}^j| \rceil$ bits[3] to store it. Hence, $\mathbf{E}_C^j = \lceil \log |\mathbf{U}_\mathbf{M}^j| \rceil$.

**Input(s) and Output(s) of the Algorithm:**

For each column of the table, the merging algorithm combines the main and delta partitions of the column into a single (modified) main partition and creates a new empty delta partition. In addition, the dictionary $\mathbf{U}'^j$ maintained for each column of the main table is also updated to reflect the modified merged column. This also includes modifying the compressed values stored for the various tuples in the merged column.

For the $j^{th}$ column, the input for the merging algorithm consists of $\mathbf{M}^j$, $\mathbf{D}^j$ and $\mathbf{U}_\mathbf{M}^j$, while the output consists of $\mathbf{M}'^j$ and $\mathbf{U}_\mathbf{M}'^j$. Furthermore, we define the cardinality $\mathbf{N}'_M$ of the output and the size of the merged dictionary $|\mathbf{U}_\mathbf{M}'^j|$ as shown in Equation 2 and 3.

$$\mathbf{N}'_M = \mathbf{N}_M + \mathbf{N}_D \qquad (2)$$
$$|\mathbf{U}_\mathbf{M}'^j| = |\mathbf{D}^j \cup \mathbf{U}_\mathbf{M}^j| \qquad (3)$$

---
[3]Unless otherwise stated, log refers to logarithm with base 2 ($\log_2$).

We perform the merge using the following two steps:

(1) **Merging Dictionaries:** This step consists of the following two sub-steps: a) Extracting the unique values from the delta partition $\mathbf{D}^j$ to form the corresponding sorted dictionary denoted as $\mathbf{U}_\mathbf{D}^j$. b) Merging the two sorted dictionaries $\mathbf{U}_\mathbf{M}^j$ and $\mathbf{U}_\mathbf{D}^j$, creating the sorted dictionary $\mathbf{U}_\mathbf{M}'^j$ without duplicate values.

(2) **Updating Compressed Values:** This step consists of appending the delta partition $\mathbf{D}^j$ to the main partition $\mathbf{M}^j$ and updating the compressed values for the tuples, based on the new dictionary $\mathbf{U}_\mathbf{M}'^j$. Since $\mathbf{D}^j$ may have introduced new values, this step requires: a) Computing the new compressed value-length. b) Updating the compressed values for all tuples with the new compressed value, using the index of the corresponding uncompressed value in the new dictionary $\mathbf{U}_\mathbf{M}'^j$.

We now describe the above two steps in detail and also compute the order of complexity for each of the steps. As mentioned earlier, the merging algorithm is executed separately for each column of the table.

### 5.1 Merging Dictionaries (Step 1)

The basic outline of step one is similar to the algorithm of a sort-merge-join[20]. However, instead of producing pairs as an output of the equality comparison, the merge will only generate a list of unique values. The merging of the dictionaries is performed in two steps (a) and (b).

*Step 1 (a).* This step involves building the dictionary for the delta partition $\mathbf{D}^j$. Since we maintain a CSB+ tree to support efficient insertions into $\mathbf{D}^j$, extracting the unique values in a sorted order involves a linear traversal of the leaves of the underlying tree structure [24]. The output of Step 1(a) is a sorted dictionary for the delta partition $\mathbf{U}_\mathbf{D}^j$, with a run-time complexity of $\mathcal{O}(|\mathbf{U}_\mathbf{D}^j|)$

*Step 1(b).* This step involves a linear traversal of the two sorted dictionaries $\mathbf{U}_\mathbf{M}^j$ and $\mathbf{U}_\mathbf{D}^j$ to produce a sorted dictionary $\mathbf{U}_\mathbf{M}'^j$. In line with a usual merge operation, we maintain two pointers called $iterator_\mathcal{M}$ and $iterator_\mathcal{D}$, to point to the values being compared in the two dictionaries $\mathbf{U}_\mathbf{D}^j$ and $\mathbf{U}_\mathbf{M}^j$, respectively. Both are initialized to the start of their respective dictionaries. At each step, we compare the current values being pointed to and append the smaller value to the output. Furthermore, the pointer with the smaller value is also incremented. This process is carried out till the end of one of the dictionaries is reached, after which the remaining dictionary values from the other dictionary are appended to the output dictionary. In case both values are identical the value is appended to the dictionary once and the pointers for the dictionaries are incremented. The output of Step 1(b) is a sorted dictionary $\mathbf{U}_\mathbf{M}'^j$ for the merged column, with $|\mathbf{U}_\mathbf{M}'^j|$ denoting its cardinality. The run-time complexity of this step is $\mathcal{O}(|\mathbf{U}_\mathbf{M}^j| + |\mathbf{U}_\mathbf{D}^j|)$.

### 5.2 Updating Compressed Values (Step 2)

The compressed values are updated in two steps – (a) computing the new compressed value length and (b) writing the new main partition and updating the compressed values.

*Step 2(a).* The new compressed value-length is computed as shown in Equation 4. Note that the length for storing the compressed values in the column may has increased from the one used for storing the compressed values before the merging algorithm. Since we use the same length for all the compressed values, this step executes in $\mathcal{O}(1)$ time.

$$\mathbf{E}_C'^j = \lceil \log(|\mathbf{U}_\mathbf{M}'^j|) \rceil \text{ bits} \qquad (4)$$

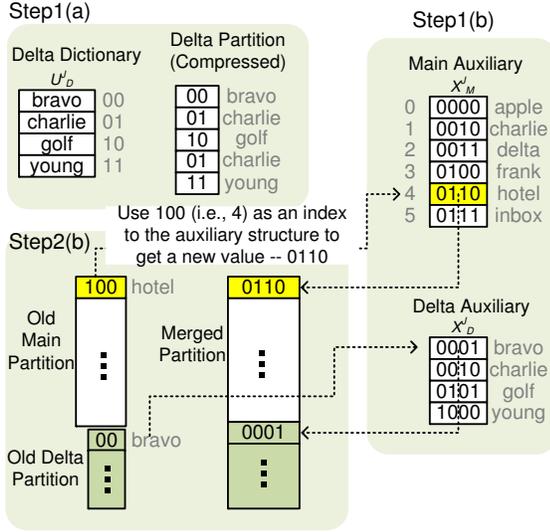

Figure 6: Example showing the various steps executed by our linear-time merging algorithm. The values in the column are similar to those used in Figure 5.

*Step 2(b).* We need to append the delta partition to the main partition and update the compressed values. As far as the main partition $\mathbf{M}^j$ is concerned, we use the following methodology. We iterate over the compressed values and for a given compressed value $K_C^i$, we compute the corresponding uncompressed value $K^i$ by performing a lookup in the dictionary $\mathbf{U}_\mathbf{M}^j$. We then search for $K^i$ in the updated dictionary $\mathbf{U}_\mathbf{M}'^j$ and store the resultant index as the encoded value using the appropriate number of $\mathbf{E}_C^j$ bits in the output. For the delta partition $\mathbf{D}^j$, we already store the uncompressed value, hence it requires a search in the updated dictionary to compute the index, which is then stored.

Since the dictionary is sorted on the values, we use a binary search algorithm to search for a given uncompressed value. The resultant run-time of the algorithm is

$$\mathcal{O}(\mathbf{N}_M + (\mathbf{N}_M + \mathbf{N}_D) \cdot \log(|\mathbf{U}_\mathbf{M}'^j|)). \quad (5)$$

To summarize the above, the total run-time for the merging algorithm is dominated by Step 2(b) and heavily dependent on the search run-time. As shown in Section 7, this makes the merging algorithm prohibitively slow and infeasible for current configurations of tables. We now present an efficient variant of Step 2(b), which performs the search in linear time at the expense of using an auxiliary data structure. Since merging is performed on every column separately, we expect the overhead of storing the auxiliary structure to be very small as compared to the total storage and independent of the number of columns in a table and the number of tables residing in the main memory.

### 5.3 Initial Performance Improvements

Based on the previously described naïve algorithm, we significantly increase the merge performance by adding an additional auxiliary data structure per main and per delta partition, denoted as $\mathbf{X}_\mathbf{M}^j$ and $\mathbf{X}_\mathbf{D}^j$ respectively. The reasoning for the auxiliary data structure is to provide a translation table with constant access cost during the expensive Step 2(b). We now describe the modified Steps 1(a), 1(b) and 2(b) to linearize the update algorithm of compressed values and improve the overall performance.

*Modified Step 1(a).* In addition to computing the sorted dictionary for the delta partition, we also replace the uncompressed values in the delta partition with their respective indices in the dictionary. By this approach lookup indices for Step 2 are changed to fixed width and allow better utilization of cache lines and CPU architecture aware optimizations like SSE.

Since our CSB+ tree structure for the delta partition also maintains a list of tuple ids with each value, we access these values while performing the traversal of the tree leaves and replace them by their newly computed index into the sorted dictionary. Although this involves non-contiguous access of the delta partition, each tuple is only accessed once, hence the run-time is $\mathcal{O}(\mathbf{N}_D)$. For example, consider Figure 6, borrowing the main/delta partition values depicted in Figure 5. As shown in Step 1(a), we create the dictionary for the delta partition (with 4 distinct values) and compute the compressed delta partition using 2 bits to store each compressed value.

*Modified Step 1(b).* In addition to appending the smaller value (of the two input dictionaries) to $\mathbf{U}_\mathbf{M}'^j$, we also maintain the index to which the value is written. This index is used to incrementally map each value from $\mathbf{U}_\mathbf{D}^j$ and $\mathbf{U}_\mathbf{M}^j$ to $\mathbf{U}_\mathbf{M}'^j$ in the selected mapping table $\mathbf{X}_\mathbf{M}^j$ or $\mathbf{X}_\mathbf{D}^j$. If both compared values are equal the same index will be added to the two mapping tables.

At the end of Step 1(b), each entry in $\mathbf{X}_\mathbf{M}^j$ corresponds to the location of the corresponding uncompressed value of $\mathbf{U}_\mathbf{M}^j$ in the updated $\mathbf{U}_\mathbf{M}'^j$. Similar observations hold true for $\mathbf{X}_\mathbf{D}^j$ (w.r.t. $\mathbf{U}_\mathbf{D}^j$). Since this modification is performed while building the new dictionary and both $\mathbf{X}_\mathbf{M}^j$ and $\mathbf{X}_\mathbf{D}^j$ are accessed in a sequential fashion while populating them, the overall run-time of Step 1(b) remains as noted in Equation 3 – linear in sum of number of entries in the two dictionaries. Step 1(b) in Figure 6 depicts the corresponding auxiliary structure for the example in Figure 5.

*Modified Step 2(b).* In contrast to the original implementation described earlier, computing the new compressed value for the main (or delta) table reduces to reading the old compressed value $\mathbf{K}_C^i$ and retrieving the value stored at $\mathbf{K}_C^i th$ index of the corresponding auxiliary structure $\mathbf{X}_\mathbf{M}^j$ or $\mathbf{X}_\mathbf{D}^j$. For example in Figure 6, the first compressed value in the main partition has a compressed value of 4 ($100_2$ in binary).

In order to compute the new compressed value, we look up the value stored at index 4 in the auxiliary structure that corresponds to 6 ($1100_2$ in binary). So value 6 is stored as the new compressed value, using 4 bits since the merged dictionary has 9 unique values, as shown in Figure 5. Therefore, a lookup and binary search in the original algorithm description is replaced by a lookup in the new algorithm, reducing the run-time to $\mathcal{O}(\mathbf{N}_M + \mathbf{N}_D)$.

To summarize, the modifications described above result in a merging algorithm with overall run-time of

$$\mathcal{O}(\mathbf{N}_M + \mathbf{N}_D + |\mathbf{U}_\mathbf{M}^j| + |\mathbf{U}_\mathbf{D}^j|) \quad (6)$$

which is linear in terms of the total number of tuples and a significant improvement compared to Equation 5.

## 6. ARCHITECTURE-AWARE MERGE IMPLEMENTATION

In this section, we describe our optimized merge algorithm on modern CPUs in detail and provide an analytical model highlighting the corresponding compute and memory traffic requirements. We first describe the scalar single-threaded algorithm and later extend it to exploit the multiple cores present on current CPUs.

The model serves the following purposes: (1) Computing the efficiency of our implementation. (2) Analyzing the performance and projecting performance for varying input parameters and underly-ing architectural features like varying core and memory bandwidth.

## 6.1 Scalar Implementation

Based on the modified Step 1(a) (Section 5.3), extracting the unique values from $\mathbf{D}^j$ involves an in-order traversal of the underlying tree structure. We perform an efficient CSB+ tree traversal using the cache-friendly algorithm described by Rao et al. [24]. The number of elements in each node (of cache-line size) depends on the size of the uncompressed values $\mathbf{E}^j$ in the $j^{th}$ column of the delta partition. For example, with $\mathbf{E}^j = 16$ bytes, each node consists of a maximum of 3 values. In addition to appending the value to the dictionary during the in-order traversal, we also traverse the list of tuple-ids associated with that value and replace the tuples with the newly computed index into the sorted dictionary. Since the delta partition is not guaranteed to be cache-resident at the start of this step (irrespective of the tree sizes), the run-time of Step 1(a) depends on the available external memory bandwidth.

As far as the total amount of data read from the external memory is concerned, the total amount of memory required to store the tree is around 2X [24] the total amount of memory consumed by the values themselves. In addition to traversing the tree, writing the dictionary $\mathbf{U}_\mathbf{D}^j$ involves fetching the data for write. Therefore, the total amount of bandwidth required for the dictionary computation is around $4 \cdot \mathbf{E}^j$ bytes per value ($3 \cdot \mathbf{E}^j$ bytes read and $1 \cdot \mathbf{E}^j$ bytes written) for the column. Updating the tuples involves reading their tuple id and a random access into $\mathbf{D}^j$ to update the tuple. Since each access would read a cache-line ($\mathbf{L}$ bytes wide) the total amount of bandwidth required would be $(2 \cdot \mathbf{L} + 4)$ bytes per tuple (including the read for the write component). This results in the total required memory traffic for this operation as shown by Equation 8. Note that at the end of Step 1(a), $\mathbf{D}^j$ also consists of compressed values (based on its own dictionary $\mathbf{U}_\mathbf{D}^j$).

Applying the modified Step 1(b) (as described in Section 5.3), the algorithm iterates over the two dictionaries and produces the output dictionary with the auxiliary structures. As far as the number of operations is concerned, each element appended to the output dictionary involves around 12 ops[4] [5]. As far as the required memory traffic is concerned, both $\mathbf{U}_\mathbf{M}^j$ and $\mathbf{U}_\mathbf{D}^j$ are read sequentially, while $\mathbf{U}_\mathbf{M}^{\prime j}$, $\mathbf{X}_\mathbf{M}^j$ and $\mathbf{X}_\mathbf{D}^j$ are written in a sequential order. Note that the compressed value-length used for each entry in the auxiliary structures is

$$\mathbf{E}_C^{\prime j} = \lceil \log(|\mathbf{U}_\mathbf{M}^{\prime j}|) \rceil. \tag{7}$$

Hence, the total amount of required read memory traffic can be calculated as shown in Equation 9. The required write memory traffic for building the new dictionary and generate the translation table is calculated as shown in Equation 10.

$$\begin{aligned}
MT &= 4 \cdot \mathbf{E}^j \cdot |\mathbf{U}_\mathbf{D}^j| + (2 \cdot \mathbf{L} + 4) \cdot \mathbf{N}_D & (8) \\
MT &= \mathbf{E}^j \cdot (|\mathbf{U}_\mathbf{M}^j| + |\mathbf{U}_\mathbf{D}^j| + |\mathbf{U}_\mathbf{M}^{\prime j}|) + \\
&\quad \frac{\mathbf{E}_C^{\prime j} \cdot (|\mathbf{X}_\mathbf{M}^j| + |\mathbf{X}_\mathbf{D}^j|)}{8} & (9) \\
MT &= \mathbf{E}^j \cdot |\mathbf{U}_\mathbf{M}^{\prime j}| + \frac{\mathbf{E}_C^{\prime j} \cdot (|\mathbf{X}_\mathbf{M}^j| + |\mathbf{X}_\mathbf{D}^j|)}{8} & (10)
\end{aligned}$$

As shown for the modified Step 2(b) in Section 5.3, the algorithm iterates over the compressed values in $\mathbf{M}^j$ and $\mathbf{D}^j$ to produce the compressed values in the output table $\mathbf{M}^{\prime j}$. For each compressed input value, the new compressed value is computed using a lookup into the auxiliary structure $\mathbf{X}_\mathbf{M}^j$ for the main partition or $\mathbf{X}_\mathbf{D}^j$ for the delta partition, with an offset equal to the stored compressed value itself.

---
[4]1 op implies 1 operation or 1 executed instruction.

The function shown in Equation 11 is executed for each element of the main partition and similar for the delta partition.

$$\mathbf{M}^{\prime j}[i] \leftarrow \mathbf{X}_\mathbf{M}^j[\mathbf{M}[i]] \tag{11}$$

As far as the memory access pattern is concerned, updating each successive element in the main or delta partition may access a random location in the auxiliary data structure (depending on the stored compressed value). Since there may not exist any coherence in the values stored in consecutive locations, each access can potentially access a different cache line (size $\mathbf{L}$ bytes). For scenarios where the complete auxiliary structure cannot fit in the on-die caches, the amount of read memory traffic to access the auxiliary data structure is approximated by Equation 12.

$$MT = \mathbf{L} \cdot (\mathbf{N}_M + \mathbf{N}_D) \tag{12}$$

In addition, reading the main/delta partition requires a read memory traffic as shown in Equation 13, while writing out the concatenated output column requires a total memory traffic that can be calculated as in Equation 14.

$$MT = \mathbf{E}_C^j \cdot (\mathbf{N}_M + \mathbf{N}_D)/8 \tag{13}$$

$$MT = 2\mathbf{E}_C^{\prime j}(\mathbf{N}_M + \mathbf{N}_D)/8 \tag{14}$$

In case $\mathbf{X}_\mathbf{M}^j$ (or $\mathbf{X}_\mathbf{D}^j$) fits in the on-die caches, their access will be bound by the computation rate of the processor, and only the main/delta partitions will be streamed in and the concatenated table is written (streamed) out. As far as the relative time spent in each of these steps is concerned, Step 1 takes about 33% of the total merge time (with $\mathbf{E}^j = 8$ bytes and 50% unique values) and Step 2 takes the remainder.[5] In terms of evidence of compute and bandwidth bound, our analytical model defines upper bounds on the performance, if the implementation was indeed bandwidth bound (and a different bound if compute bound). Our experimental evaluations show that our performance closely matches the lower of these upper bounds for compute and bandwidth resources — which proves that our performance is bound by the resources as predicted by the model.

## 6.2 Exploiting Thread-level Parallelism

We now present the algorithms for exploiting the multiple cores/threads available on modern CPUs. $\mathbf{N}_T$ denotes the number of available processing threads.

### 6.2.1 Parallelization of Step 1

Recalling Step 1(a), we perform an in-order traversal of the CSB+ tree and simultaneously update the tuples in the delta partition with the newly assigned compressed values.

There exist two different strategies for parallelization:

(i) Dividing the columns within a table amongst the available threads: Since the time spent on each column varies based on the number of unique values, dividing the columns evenly amongst threads may lead to load imbalance between threads. Therefore we use a task queue based parallelization scheme [1] and enqueue each column as a separate task. If the number of tasks is much larger than the number of threads (as in our case with few tens to hundred columns and few threads), the task queue mechanism of migrating tasks between threads works well in practice to achieve a good load balance.

---
[5]Section 7 gives more details.

(ii) Parallelizing the execution of Step 1(a) on each column amongst the available threads: Since a small portion of the run-time is spent in computing the dictionary, we execute it on a single-thread, and keep a cumulative count of the number of tuples that needs to be updated as we create the dictionary. We parallelize the next phase where these tuples are evenly divided amongst the threads and each thread scatters the compressed values to the delta partition.

For a table with very few columns, scheme (ii) performs better than scheme (i). We implemented both (i) and (ii) and since our input table consisted of few tens to hundreds of columns, we achieved similar scaling for both these schemes on current CPUs. In Section 7.2, we report the results for (i) – the results for (ii) would be similar. Step 1(b) involves merging the two sorted dictionaries $\mathbf{U}_\mathbf{M}^j$ and $\mathbf{U}_\mathbf{D}^j$ with duplicate removal and simultaneously populating the auxiliary structures $\mathbf{X}_\mathbf{M}^j$ and $\mathbf{X}_\mathbf{D}^j$. This is an inherent sequential dependency in this merge process and also requires the merge process to remove duplicates. Also note that for tables with large fractions of unique values or large value-lengths ($\geq 8$ bytes) a significant portion of the total run-time is spent in Step 1(b). Therefore, it is imperative to parallelize well, in order to achieve speedups in the overall run-time of the merging algorithm. We now describe our parallelization scheme in detail that in practice achieves a good load-balance. Let us first consider the problem of parallelizing the merging of $\mathbf{U}_\mathbf{M}^j$ and $\mathbf{U}_\mathbf{D}^j$ without duplicate removal. In order to evenly distribute the work among the $\mathbf{N}_T$ threads it is required to partition both dictionaries into $\mathbf{N}_T$-quantiles. Since both dictionaries are sorted this can be achieved in $\mathbf{N}_T \log(|\mathbf{U}_\mathbf{M}^j| + |\mathbf{U}_\mathbf{D}^j|)$ steps [8]. Furthermore, we can also compute the indices in the two dictionaries for the $i^{th}$ thread $\forall i \in \mathbf{N}_T$ following the same algorithm as presented in [5]. Thus, each thread can compute its start and end indices in the two dictionaries and proceed with the merge operation. In order to handle duplicate removal while merging, we use the following technique consisting of three phases. We additionally maintain an array *counter* of size $(\mathbf{N}_T + 1)$ elements.

*Phase 1.* Each thread computes its start and end indices in the two dictionaries and writes the merged output, while locally removing duplicates. Since the two dictionaries consisted of unique elements to start with, the only case where this can create duplicate elements is when the last element produced by the previous thread matches the first element produced by the current thread. This case is checked for by comparing the start elements in the two dictionaries with the previous elements in the respectively other dictionary. In case there is a match, the corresponding pointer is incremented before starting the merge process. Once a thread (say the $i^{th}$ thread) completes the merge execution, it stores the number of unique elements produced by that thread to the corresponding location in the counter array (i.e. *counter*[$i$]). There is an explicit global barrier at the end of phase 1.

*Phase 2.* In the second phase, we compute the prefix sum of the counter array, so that *counter*[$i$] corresponds to the total number of unique values that would be produced by the previous $i$ threads. Additionally, *counter*[$\mathbf{N}_T$] is the total number of unique values that the merge operation would produce. We parallelize the prefix sum computation using the algorithm by Hillis et al. [12].

*Phase 3.* The counter array produced at the end of phase 2 also provides the starting index at which a thread should start writing the locally computed merged dictionary. Similar to phase 1, we recompute the start and end indices in the two dictionaries. Now consider the main partition. The range of indices computed by the thread for $\mathbf{U}_\mathbf{M}^j$ also corresponds to the range of indices for which the thread can populate $\mathbf{X}_\mathbf{M}^j$ with the new indices for those values. Similar observations hold for the delta partition. Each thread performs the merge operation within the computed range of indices to produce the final merged dictionary and auxiliary data structures.

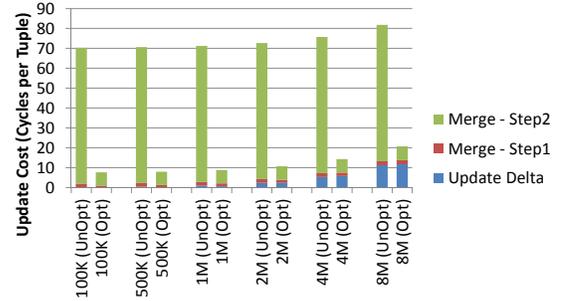

Figure 7: Update Costs for Various Delta Partition Sizes with a main partition size of 100 million tuples with 10% unique values using 8-byte values. Both optimized (Opt) and unoptimized (UnOpt) merge implementations were parallelized.

*Summary.* In comparison to the single-threaded implementation, the parallel implementation reads the dictionaries twice and also writes the output dictionary one additional time, thereby increasing the total memory traffic by

$$\mathbf{E}^j \cdot \left(|\mathbf{U}_\mathbf{M}^j| + |\mathbf{U}_\mathbf{D}^j|\right) + 2\mathbf{E}^j \cdot |\mathbf{U}_\mathbf{M}'^j| \qquad (15)$$

bytes. The overhead of the start and end index computation is also very small as compared to the total computation performed by Step 1(b). The resultant parallel algorithm evenly distributes the total amount of data read/written to each thread, thereby completely exploiting the available memory bandwidth.

### 6.2.2 Parallelization of Step 2

To parallelize the updating of compressed values, we evenly divide the total number of all tuples $\mathbf{N}_M'$ amongst the available threads. Specifically, each thread is assigned $\mathbf{N}_M'/\mathbf{N}_T$ tuples to operate upon. Since each thread reads/writes from/to independent chunks of tables, this parallelization approach works well in practice. Note that in case any of $\mathbf{X}_\mathbf{M}^j$ or $\mathbf{X}_\mathbf{D}^j$ can completely fit in the on-die caches, this parallelization scheme still exploits to read the new index for each tuple from the caches and that the run-time is proportional to the amount of bandwidth required to stream the input and output tables.

## 7. PERFORMANCE EVALUATION

We now evaluate the performance of our algorithm on a dual-socket six-core Intel® Xeon® processor X5680 with 2-way SMT per core, and each core operating at a frequency of 3.3 GHz. Each socket has 32 GB of DDR (for a total of 64 GB of main memory). The peak external memory bandwidth on each socket is 30 GB/sec. We used SUSE SLES 11 as operating system, the pthread library and the Intel® ICC 11.1 as compiler. As far as the input data is concerned, the number of columns in the partition $\mathbf{N}_C$ varies from 20 to 300. The value-length of the uncompressed value $\mathbf{E}^j$ for a column is fixed and chosen from 4 bytes to 16 bytes. The number of tuples in the main partition $\mathbf{N}_M$ varies from 1 million to 1 billion, while the number of tuples in the delta partition $\mathbf{N}_D$ varies from 500,000 to 50 million, with a maximum of around 5% of $\mathbf{N}_M$. Since the focus of the paper is on in-memory databases, the number of columns is chosen so that the overall data completely fits in the available main memory of the CPU. The fraction of unique values $\lambda_M^j$ and $\lambda_D^j$ varies from 0.1% to 100% to cover the spectrum of scenarios in real applications (described in Section 2). For all experiments, the values are generated uniformly at random. We chose uniform value distributions, as this represents the worst possible cache utilization for the values and auxiliary structures. Different value distributions can only improve cache utilization, leading to better merge times. However, differences in merge times due to different value distributions are expected to be very small and are therefore neglected. We first show the impact of varying $\mathbf{N}_D$ on the merge performance. We then vary $\mathbf{E}^j$ from 4-16 bytes to analyze the effect of varying value-lengths on merge operations. We finally vary the percentage of unique values ($\lambda_M^j$, $\lambda_D^j$) and the size of the main partition $\mathbf{N}_M$. In order to normalize performance w.r.t. varying input parameters, we introduce the term – *update cost*. Update Cost is defined as the amortized time taken per tuple per column (in cycles/tuple), where the total time is the sum of times taken to update the delta partition $\mathbf{T}_U$ and the time to perform the merging of main and delta partitions $\mathbf{T}_M$, while the total number of tuples is $\mathbf{N}_M + \mathbf{N}_D$.

## 7.1 Impact of Delta Partition Size

Figure 7 shows the update cost for varying tuples of the delta partition. In addition to the delta partition update cost, we also show both run-times for the unoptimized and optimized Steps 1 and 2 in the graph. $\mathbf{N}_M$ is fixed to be 100 million tuples, while $\mathbf{N}_D$ is varied from 500,000 (0.5%) to 8 million (8%) tuples. $\lambda_M^j$ and $\lambda_D^j$ are fixed to be around 10%. The uncompressed value-length $\mathbf{E}^j$ is 8 bytes. We fix the number of columns $\mathbf{N}_C$ to 300. Note that the run-times are on a *parallelized code for both implementations* on our dual-socket multi-core system.

As far as the unoptimized merge algorithm is concerned, Step 2 (updating the compressed values) takes up the majority of the run-time and does not change (per tuple) with the varying number of tuples in the delta partition. The optimized Step 2 algorithm drastically reduces the time spent in the merge operation (by 9-10 times) as compared to the unoptimized algorithm. Considering the optimized code, as the delta partition size increases, the percentage of the total time spent on delta updates increases and is 30% – 55% of the total time. This signifies that the overhead of merging contributes a relatively small percentage to the run-time, thereby making our scheme of maintaining separate main and delta partitions with the optimized merge an attractive option for performing updates without a significant overhead.

The update rate in tuples/second is computed by dividing the total number of updates with the time taken to perform delta updates and merging the main and delta partitions for the $\mathbf{N}_C$ columns. As an example, for $\mathbf{N}_D$ = 4 million and say $\mathbf{N}_C$ = 300, an update cost of 13.5 cycles per tuple (from Figure 7) evaluates to

$$\frac{4,000,000 \cdot 3.3 \cdot 10^9}{13.5 \cdot 104,000,000 \cdot 300} \approx 31,350 \text{ updates/second.} \quad (16)$$

## 7.2 Impact of Value-Length and Percentage of Unique values

Figure 8 shows the impact of varying uncompressed value-lengths on the update cost. The uncompressed value-lengths are varied between 4, 8 and 16 bytes. We show two graphs with the percentage of unique values fixed at (a) 1% and (b) 100% respectively. $\mathbf{N}_M$ is fixed to be 100 million tuples for this experiment and the breakdown of update cost for $\mathbf{N}_D$ equal to 1 million and 3 million tuples is shown. We fix the number of columns $\mathbf{N}_C$ to 300.

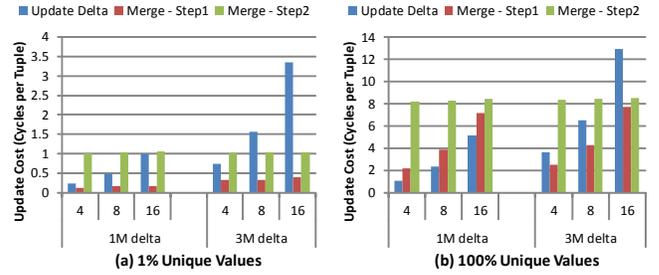

Figure 8: Update Costs for Various Value-Lengths for two delta sizes with 100 million tuples in the main partition for 1% and 100% unique values.

As the value-length increases, the time taken per tuple to update the delta partition increases and becomes a major contributor to the overall run-time. This time also increases as the size of the delta partition increases. For example, in Figure 8(a), for an uncompressed value-length of 16 bytes, the delta update time increases from about 1.0 cycles per tuple for $\mathbf{N}_D$ = 1 million to about 3.3 cycles for $\mathbf{N}_D$ = 3 million. This time increases as the percentage of unique values increases. The corresponding numbers in Figure 8(b) for 100% unique values are 5.1 cycles for $\mathbf{N}_D$ = 1 million and 12.9 cycles for $\mathbf{N}_D$ = 3 million.

As far as the Step 2 of the merge is concerned, the run-time depends on the percentage of unique values. For 1% unique values, the auxiliary structures fit in cache. As described in Section 6.2, the auxiliary structures being gathered fit in cache and the run-time is bound by the time required to read the input partitions and write the updated partitions. We get a run-time of 1.0 cycles per tuple (around 1.8 cycles per tuple on 1-socket), which is close to the bandwidth bound computed in Section 6.2. For 100% unique values, the auxiliary structures do not fit in cache and must be gathered from memory. The time taken is then around 8.3 cycles (15.0 cycles on 1-socket), which closely matches (within 10%) the analytical model developed in Section 6. The time for Step 2 mainly depends on whether the auxiliary structures can fit in cache and therefore is constant with small increases in the delta size from 1 million to 3 million.

As far as Step 1 is concerned, for a given delta partition size $\mathbf{N}_D$, the time spent in Step 1 increases sub-linearly with the increase in value-length (Section 6.1). For a fixed value-length, the time spent increases marginally with the increase in $\mathbf{N}_D$ – due to the fact that this increase in partition size only changes the unique values by a small amount and hence the compressed value-length also changes slightly, resulting in a small change in the run-time of Step 1. With larger changes in the percentage of unique values from 1% to 100%, the run-time increases. For instance, for 8-byte values and 1 million delta partitions, Step 1 time increases from 0.1 cycles per tuple at 1% unique values to 3.3 cycles per tuple at 100% unique values.

Finally, the percentage of time spent in updating the tuples as compared to the total time increases both with increasing value-lengths for fixed $\mathbf{N}_D$ and increase in $\mathbf{N}_D$ for fixed value-lengths.

*Parallel Scalability.* We now highlight the parallel scalability of our system for two cases: (a) 1% unique values and (b) 100% unique values. We report the execution times on 1-socket, with the respective 2-socket scaling reported in a separate column. We achieve near-linear scaling with 2-sockets (1.8 – 2.0 times) for various phases of our algorithm. This is in-line with the 2 times peak increase in the computing resources from 1 to 2 sockets.

The update costs in cycles per tuple for serial (1 thread) and par-allel (6 threads) for 1 million delta size and 8-byte values are shown in Table 2. All performance run-times below are for a single-socket execution, to explain the thread scaling per socket. We first note that in all cases, updating the delta partitions takes place when a tuple is updated in the database and not during the merge. When a tuple is updated, multiple delta partitions corresponding to different columns of the tuple need to be modified at the same time. Consequently, in this phase, we parallelize over the different columns being updated. This scheme gives us a parallel scalability of about 5 times.

| % unique | Step | Update cost (cpt) 1T | Update cost (cpt) 6T | Scaling on 1-socket | Socket Scaling |
|---|---|---|---|---|---|
| 1% | Update Delta | 4.52 | 0.87 | 5.2X | 1.8X |
|  | Step 1 | 1.29 | 0.30 | 4.3X | 1.9X |
|  | Step 2 | 3.89 | 1.85 | 2.1X | 1.9X |
| 100% | Update Delta | 20.63 | 4.21 | 4.9X | 1.9X |
|  | Step 1 | 20.92 | 6.97 | 3.0X | 2.0X |
|  | Step 2 | 66.21 | 15.0 | 4.4X | 1.8X |

Table 2: Parallel scalability of various steps for different percentages of unique values. 1T denotes single-threaded run, while 6T represents the run using all 6-cores on a single socket. Socket Scaling denotes the additional scaling using both sockets on our system as compared to a single-socket execution.

During merge time, Steps 1 and 2 of the merge are parallelized as described in Section 6.2 for each partition. For the case when only 1% of the partition values are unique, Step 1 is compute bound. However, the parallel Step 1 algorithm, as described in Section 6.2, performs twice as many comparisons as the serial code due to the three phase algorithm. This parallel overhead results in a scaling of 4.3 times rather than 6 times on our 6-core system. The parallel Step 2, for 1% unique values is bound by memory bandwidth. This is the streaming bandwidth involved in bringing in the partitions from memory, while the auxiliary structures are in cache. Therefore, Step 2 does not scale linearly with compute resources and only achieves 2.1 times scaling on 6-cores.

As far as 100% unique values are concerned, all data structures must be accessed from memory. The parallel Steps 1 and 2 are both bandwidth bound. While Step 1 involves streaming accesses from memory, Step 2 involves irregular accesses (gathers) of auxiliary structures from memory. In terms of the serial single-threaded code, Step 1 has very regular accesses for which the hardware prefetcher is successful at prefetching data and hence serial performance is good. This results in a low speedup of 3 times for serial code. Step 2 is also bandwidth bound and furthermore the parallel code accesses more data than the serial code. However, due to irregular memory accesses, the hardware prefetcher does not work well and the serial code is heavily latency bound. On the other hand, the parallel version introduces more parallel memory accesses which reduce latency impacts. We get better parallel scaling of 4.4 times for Step 2 than the 3 times for Step 1. Also note that the Step 2 scaling for 100% unique values is better than for 1% unique values since the serial code gets better performance when data starts fitting in caches.

Finally, we note that the use of Simultaneous Multithreading (SMT) does not improve performance. As mentioned above this is because our optimized implementations are either compute or bandwidth bound while SMT mainly helps latency bound applications.

### 7.3 Impact of Main Partition Size and Percentage of Unique Entries

Figure 9 shows the impact of varying the percentage of unique entries and the main partition size on the update performance (higher numbers are better). In this figure, the uncompressed value-length is fixed to the common practical scenario of 8 bytes. The size of the delta partition is fixed at 1% of the main partition size in this experiment. The update cost is shown as the percentage of unique entries varies from 0.1% to 100% to cover a wide range of practical scenarios. We also vary the main partition size from 1 million to 1 billion. We fix the number of columns $N_C$ to 300.

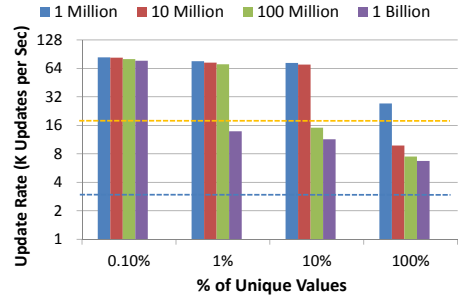

Figure 9: Update Rates for optimized merge with varying main partition sizes (1 million to 1 billion tuples) and varying percentage of unique values (0.1% to 100%). The delta partition size is fixed at 1% of the main partition. The two dashed lines show our low and high target update rates of 3,000 and 18,000 updates/second respectively.

Figure 9 clearly shows the impact of caching on the overall update performance. The cost of Step 2, which is a significant fraction of overall update cost, depends on the dictionary size of the main and delta auxiliary structures. When both main and delta auxiliary structures $X_M^j$ and $X_D^j$ fit in cache, Step 2 operates out of cache and is not limited by main memory bandwidth but rather from CPU compute resources. This leads to a high performance of over 81,000 updates/second. When the auxiliary structures do not fit in cache, the update rate is limited by memory bandwidth to about 7,100 updates/second, also matching our model. The number of entries in the auxiliary structures is proportional to the main partition size as well as the percentage of unique entries. From the figure (for instance at 1% unique entries), there is a sharp performance difference when the main partition size changes from 100 million to 1 billion – this corresponds to the auxiliary structures size ranging from about 1 million (2.5 MB) to 10 million (30 MB). The actual cache size on our dual-socket platform is 24 MB, matching the drop in performance.

Even in cases where the auxiliary structures fit in cache, the update rate drops slightly with increasing main partition sizes. Although Step 2 time is relatively constant, the cost of inserting elements into the CSB+ tree and the cost of constructing the auxiliary structures (Step 1) increase. This can be seen for both 0.1% and 1% unique values in Figure 9.

When the auxiliary structures do not fit in cache, the update rates slightly drop with increasing main partition sizes, but stabilize at about 7,100 updates/second. This exceeds our target of 3,000 updates per second. In particular, this occurs even for very large partitions of 1 billion entries, showing that our system is scalable to future databases with larger partitions.

In Section 4, we described two target update rates – 18,000 updates/second for high update-rate systems and 3,000 for low update-rate systems. Figure 9 shows that we meet our high update rates for tables which have at most 100 million rows, when the unique values are less than 1%. We always meet our low update rates even at 100% unique values. Finally, from Figure 7, we note that the unoptimized parallelized CPU code is about 9-10 times worse than optimized code – hence we only achieve about 730 updates/second, which is much lower than our target rates.

## 7.4 Comparison With Analytical Model

We now compute the run-times projected by our analytical model and compare them to the actual obtained performance. We focus on a single-socket execution for the analysis below [6]. Let $\mathbf{N}_M$ be 100 million tuples, and $\mathbf{N}_D$ be 1 million tuples (the scenario in Table 2). $\mathbf{E}^j$ equals 8 bytes. For streaming accesses, our system obtains a memory bandwidth of around 23 GB/sec (around 7 bytes/cycle), while random accesses result in a memory bandwidth of around 5 bytes/cycle – both measured using separate micro-benchmarks, each running with 6 threads respectively. We use these bandwidth numbers in our analysis below.

Consider the case with 100% unique entries. As such, the auxiliary structures ($\mathbf{X}_M^j$ and $\mathbf{X}_D^j$) cannot fit in the LLC of our system and the performance would be bound by the underlying memory bandwidth. Consider Step 1, consisting of steps 1(a) and 1(b). The total amount of memory traffic generated by this step is given by Equation 8. The first part consists of streaming access, and can be accessed at 7 bytes/cycle, while the second part (random access) will be at 5 bytes/cycle. The net performance equals

$$\frac{\frac{4 \cdot 8 \cdot 1,000,000}{7} + \frac{(128+4) \cdot 1,000,000}{5}}{101,000,000} = 0.306 \text{ cpt.} \quad (17)$$

The total amount of memory traffic generated in step 1(b) is obtained by adding up Equations 9, 10, and 15, to obtain around 6.6 cpt. The net time taken for Step 1 equals $0.3 + 6.6 = 6.9$ cycles, which is within 1% of the obtained performance (Table 2). Similarly for Step 2(b), the total memory traffic is obtained by adding Equation 12 (at 5 bytes/cycle) and Equation 13, 14 at 7 bytes/cycle to obtain 14.2 cycles. The obtained performance is around 15 cpt, which is within 5.5% of the projected performance by our model.

Similarly, we can compute the performance numbers for cases where the auxiliary data structures are expected to fit in the LLC. Consider 1% unique values, and Step 2. The cost of accessing the auxiliary structures will now be bound by the number of executed instructions while the remaining time (Equation 13 and 14) will still be bound by the available memory bandwidth. Assuming linear scaling of the compute-bound part, the total runtime would be around

$$\frac{4}{6} + \frac{\frac{19.9}{8} + \frac{2 \cdot 19.9}{8}}{7} = 1.73 \text{ cycles} \quad (18)$$

The actual achieved performance is 1.85 cycles (Table 2), which is within 7% performance. Our implementation run-time closely matches the projected performance of our analytical model (within 1-10%). Our model can be used to project performance with varying input scenarios. For e.g., with 0.1% unique entries, the run-times would be similar to the case shown above where $\mathbf{X}_M^j$ and $\mathbf{X}_D^j$ fit in caches. As the number of unique entries increases, $\mathbf{X}_M^j$ and $\mathbf{X}_D^j$ may cease to fit in caches and will therefore obtain performances completely bound by the memory bandwidth, with the performance numbers obtained by plugging in the appropriate parameter values of $\lambda_M^j$, $\lambda_D^j$ and $\mathbf{E}_C'^j$.

## 8. RELATED WORK

Vertical partitioned databases as HYRISE have been researched from the very first conferences on database systems with focus on read-intensive environments [26, 21, 2]. Pure vertical partitioning into a "column-store" has been a recent topic of interest in literature. Copeland and Khoshafian [7] introduced the concept of a Decomposition Storage Model (DSM) as a complete vertical and attribute-wise partitioned schema, which has been the foundation for multiple commercial and non-commercial column store implementations such as MonetDB/X100 [4], C-Store [25] or Sybase IQ [9]. All of those examples have shown the ability to outperform conventional databases in read-mostly analytical-style scenarios with high selectivity. Unlike HYRISE, most of the column-store implementations are pure disk based approaches and focus on improving the overall performance by reducing the number of disk seeks by decomposing relations. Consequently, data modifications must be propagated to multiple files on disk. This implementation is therefore inappropriate for workloads combining transactional- and analytical-style queries, because updates and inserts are spread across different disk locations. As in HYRISE, data compression can limit the applicability to scenarios with frequent updates, leading to dedicated delta structures in order to improve performance of inserts, modifications and deletes. The authors of [4] and [23] describe a concept of treating vertical fragments as immutable objects, using a separate list for deleted tuples and uncompressed delta columns for appended data while using a combination of both for updates. In contrast, HYRISE maintains all data modification of a table in one differential delta buffer. The authors of [11] also use a dedicated differential buffer in order to improve update performance in a column store. They describe a technique to merge a stable (persisted) storage with a dedicated differential buffer by (implicit) positions rather than key values. However, the focus lies on merging the differences at query processing time. They use the term *checkpointing* to describe the inevitable process of incorporating and persisting the differential buffer into the stable storage. However, they do not describe how this process works in detail and how it affects update rates when performed online. In [3] Beier et al. describe an online process for reorganization of compressed main memory databases. Even though the work is close in spirit, the focus is on re-compression of individual horizontal chunks of a table compared to our full table approach. Furthermore they deliberately focus on analytical workloads compared to the mixed workload we describe. In [16] we presented an early version of our approach for optimizing write performance in read-optimized column stores, however we did not focus on hardware optimizations and parallelism.

---

[6] Our efficiently parallelized implementation achieves near-linear scaling with two sockets: per-phase scaling presented in Table 2.

Recently, efforts have been made to utilize the high compute and bandwidth resources of modern processors to speed up database primitives such as search, scan, sort, join and aggregation [14, 27, 5, 15, 6]. Such works have focused on improving performance of specific primitives, but the impact of these optimizations on real workloads has not been described. In contrast to previous work, we start with characterizations of real world enterprise workloads, while our design decisions and experimental parameters are driven by real issues that make updates on read-optimized databases prohibitively slow.

## 9. DISCUSSION AND FUTURE WORK

In earlier sections, we presented our characterizations of enterprise workloads. This showed that apart from the traditional classification of databases into OLTP and OLAP systems single applications exist which do not fit entirely into one category. These applications like Demand Planning or ATP introduce a mixed workload, working on transactional data including write operations as well as complex read operations on large data sets. The focus of this paper is an optimized merge process, enabling read-optimized in-memory column stores to meet both requirements for read and up-

71

date performance of today's enterprise applications. The introduction of mixed workload systems meets the increasing demand for "real-time analytics", reporting directly on the transactional data. We see the potential for transactional applications using analytical functionality and for faster and more detailed reports in real time. While this trend develops, complex analytical queries and full table scan operations will increase in the database workload, justifying the decision for a read-optimized database. The rethinking of how persistence should be managed to leverage new hardware possibilities and discarding parts of the over 20-year old data management infrastructure can simplify future ERP systems. This would end the trend of more and more complex systems compensating shortcomings of the persistence layer, e.g. by maintaining pre-computed and materialized results. The overall goal should be to define an application persistence based on data characteristics and usage patterns of the consuming applications. Thus, the observations and optimizations in this paper are closely tied to the observed enterprise applications. If the workload changes, the database engine has to adapt.

For future work we see two major directions: On the one hand we plan to extend the current analytical model with a more detailed model for scans and lookup operations [19]. On the other hand, we plan to investigate other delta partition structures to balance the insert/merge costs to achieve optimal performance. In addition, resource consumption needs to be considered. The algorithms and solutions presented throughout the paper are optimized towards optimal resource utilization. However, depending on the current system load it can be advisable to prolong the merge process in favor to increase the current insert throughput. A scheduling algorithm could constantly analyze the available bandwidth and thus adjust the degree of parallelization for the merge process. The memory consumption of the merge process has to be tackled. Possible ideas include an incremental processing of the individual attributes for the cost of adding intermediate data structures to guarantee transactional safety. Ideas from [3] could be taken further to directly include a horizontal partitioning strategy.

## 10. CONCLUSIONS

In this paper, we presented a linear-time algorithm to update the compressed main storage during the merge process, which resulted in a speedup of 30 times over current implementations. The update performance facilitated by our merge algorithm enables in-memory column stores to execute transactional enterprise workloads, while retaining their high read performance and therefore allowing new and more complex mixed transactional and analytical queries on up-to-date transactional data.